# Singular Zone in Quadrotor Yaw-Position Feedback Linearization

Zhe Shen, Takeshi Tsuchiya

zheshen@g.ecc.u-tokyo.ac.jp     The University of Tokyo

**Abstract**

It is well known that the conventional quadrotor is an under-actuated MIMO system. The number of the inputs is less than the number of the degree of freedom. One approach in controlling this non-holonomic system is feedback linearization. Commonly selection in controlled outputs can be yaw and the position [1]. It is reported that no singularity is found in its decoupling matrix (delta matrix), making it possible to apply the inversion within a wide range. However, we find the ignored singular zone within the range of interest. The unreported singular area can cause the failure in the controller design. This paper visualizes this uninvertible area and details the deduction process.

## I. Introduction

With the higher demands on the performance in conventional quadrotor trajectory tracking, the linear controller around the equilibrium state is no longer suitable. Several nonlinear controllers have been used to stabilize the quadrotor. In detail, [2]-[3] use feedback linearization equipped with a PID controller. [4] uses the backstepping method, which guarantees the stability using Lyapunov criteria. [5] developed a sliding mode controller. [6] applied MPC.

Among these widely used nonlinear controllers, the feedback linearization method is relatively special since it transfers the original nonlinear system into a linear one, which provides the possibility of utilizing the linear controllers. The advantage of the dynamic inverse method put forward in [1] has its advantage of exact linearize the model with no approximation which can introduce the bias in input-output linearization.

Although feedback linearization yields the convenience in designing the exterior loop controller, several problems may hinder its applications. One issue is that the control signal and the state variable should not hit the bound/constraints; hitting the input saturation or positive constraint bounds are strictly prohibited in the exchange to guarantee the feasibility of the linear-system-based stability analysis. To meet this requirement, [7] put forward Reference Governor.

Another potential issue is the existence of inversion; an uninvertible singular matrix can block the application of the Feedback linearization though some methods, e.g., PCH [8], dynamic approximation [9], etc. are put forward. One typical way to pick the controlled variables is the yaw ($\psi$) and the position ($x, y, z$) combination. By applying two integrators in the input signal, [1] shows their results in simulation with the conclusion that the singularity does not exist once we are in the zone of interest. That is $\phi \in \left(-\frac{\pi}{2}, \frac{\pi}{2}\right)$, $\theta \in \left(-\frac{\pi}{2}, \frac{\pi}{2}\right)$. However, we will report the extra singular zone in this paper.

Since yaw-position combination is still popular as the selected controlled variable in several recent studies. It is quite necessary to correct and report the singular area in the zone of interest. We also provided thoroughly the mathematical deduction process in this paper.

The remaining of this paper is organized as follows. In Section II, previous works on the dynamics and the feedback linearization applied to a quadrotor is introduced. Section III deduces the extra singular area within the zone of interest. Several remarks are given in Section IV. A potential approach to avoid this singular zone is put forward and simulated in Section V. Finally, we make conclusions and further discussions in Section VI.

## II. Dynamics and Feedback Linearization Method in UAV Control

This section briefs the feedback linearization control in [1]. The readers are encouraged to refer to the original manuscript [1] for detail.

The input vector is given by

$$[u_1 \quad u_2 \quad u_3 \quad u_4]^T \tag{1}$$

where $u_1$ is the total thrust generated by four propellers. $u_2$, $u_3$, and $u_4$ are the difference of the square of the two relevant angular velocities which contribute to the torque along the body-fixed $x-\text{axis}, y-\text{axis}, \text{and } z-\text{axis}$, respectively.

As plotted in [1], the virtual inputs are used

$$[\bar{u}_1 \quad \bar{u}_2 \quad \bar{u}_3 \quad \bar{u}_4]^T \tag{2}$$

where

$$u_1 = \zeta \tag{3}$$

$$\dot{\zeta} = \xi \tag{4}$$

$$\dot{\xi} = \bar{u}_1 \tag{5}$$

$$\bar{u}_2 = u_2 \tag{6}$$

$$\bar{u}_3 = u_3 \tag{7}$$

$$\bar{u}_4 = u_4 \tag{8}$$

The whole dynamic of quadrotor is then written

$$\dot{\bar{x}} = f(\bar{x}) + \sum_{i=1}^{4} g_i(\bar{x}) \cdot \bar{u}_i \tag{9}$$

where $\bar{x}$ is the augmented state vector

$$\bar{x} = \left[x, y, z, \psi, \theta, \phi, \dot{x}, \dot{y}, \dot{z}, \zeta, \xi, p, q, r\right]^T \tag{10}$$

where $x, y,$ and $z$ are the position with respect to the earth frame. $\psi, \theta,$ and $\phi$ are the attitude yaw, pitch, and roll, respectively. $p, q,$ and $r$ are the angular velocities along the body-fixed $x-$ axis, $y-$ axis, and $z-$ axis, respectively.

Define $g_i(\bar{x})$ in Equation (11)-(14)

$$g_1(\bar{x}) = [0,0,0,0,0,0,0,0,0,0,1,0,0,0]^T \tag{11}$$

$$g_2(\bar{x}) = \left[0,0,0,0,0,0,0,0,0,0,0,\frac{d}{I_x},0,0\right]^T \tag{12}$$

$$g_3(\bar{x}) = \left[0,0,0,0,0,0,0,0,0,0,0,0,\frac{d}{I_y},0\right]^T \tag{13}$$

$$g_4(\bar{x}) = \left[0,0,0,0,0,0,0,0,0,0,0,0,0,\frac{d}{I_z}\right]^T \tag{14}$$

where $d$ is the length of each arm. $I_x, I_y,$ and $I_z$ are the rotational inertia along body-fixed $x-$ axis, $y-$ axis, and $z-$ axis, respectively.

$f(\bar{x})$ is defined in Equation (15).

$$f(\bar{x}) = \begin{bmatrix} \dot{x} \\ \dot{y} \\ \dot{z} \\ \dfrac{s\phi}{c\theta} \cdot q + \dfrac{c\phi}{c\theta} \cdot r \\ c\phi \cdot q - s\phi \cdot r \\ p + s\phi \cdot t\theta \cdot q + c\phi \cdot t\theta \cdot r \\ g_1^7 \cdot \zeta \\ g_1^8 \cdot \zeta \\ g_1^9 \cdot \zeta + g \\ \xi \\ 0 \\ \dfrac{I_y - I_z}{I_x} \cdot q \cdot r \\ \dfrac{I_z - I_x}{I_y} \cdot p \cdot r \\ \dfrac{I_x - I_y}{I_z} \cdot p \cdot q \end{bmatrix} \tag{15}$$

where $s\Lambda$, $c\Lambda$, and $t\Lambda$ represent $\sin(\Lambda)$, $\cos(\Lambda)$, and $\tan(\Lambda)$, respectively. $g$ is the gravitational acceleration. $g_1^7$, $g_1^8$, and $g_1^9$ are defined in Equation (16) – (18), respectively.

$$g_1^7 = -\frac{1}{m} \cdot (c\phi \cdot c\psi \cdot s\theta + s\phi \cdot s\psi) \triangleq -\frac{1}{m} \cdot A_1 \tag{16}$$

$$g_1^8 = -\frac{1}{m} \cdot (c\phi \cdot s\psi \cdot s\theta - s\phi \cdot c\psi) \triangleq -\frac{1}{m} \cdot A_2 \tag{17}$$

$$g_1^9 = -\frac{1}{m} \cdot c\theta \cdot c\phi \triangleq -\frac{1}{m} \cdot A_3 \tag{18}$$

where $m$ is the mass of the quadrotor.

The 4 interested controlled variables ($y_1$, $y_2$, $y_3$, $y_4$) are:

$$\begin{bmatrix} y_1 \\ y_2 \\ y_3 \\ y_4 \end{bmatrix} = \begin{bmatrix} x \\ y \\ z \\ \psi \end{bmatrix} \tag{19}$$

To apply the feedback linearization, [1] calculated the higher derivative of the controlled variables and received Equation (20).

$$\begin{bmatrix} y_1^{(4)} \\ y_2^{(4)} \\ y_3^{(4)} \\ \ddot{y}_4 \end{bmatrix} = \begin{bmatrix} x^{(4)} \\ y^{(4)} \\ z^{(4)} \\ \ddot{\psi} \end{bmatrix} = \mathrm{Ma}(\bar{x}) + \Delta(\bar{x}) \cdot \begin{bmatrix} \bar{u}_1 \\ \bar{u}_2 \\ \bar{u}_3 \\ \bar{u}_4 \end{bmatrix} \qquad (20)$$

where both $\mathrm{Ma}(\bar{x})$ and $\Delta(\bar{x})$ are 4 by 4 matrix of state (10).

Once the $\Delta(\bar{x})$ is invertible, the further linearization can be further applied. It is concluded in [1] that $\Delta(\bar{x})$ is always invertible given that $\phi \in \left(-\frac{\pi}{2}, \frac{\pi}{2}\right)$, $\theta \in \left(-\frac{\pi}{2}, \frac{\pi}{2}\right)$, $\zeta \neq 0$.

However, based on our deduction, this condition does not hold within the entire space of $\phi \in \left(-\frac{\pi}{2}, \frac{\pi}{2}\right)$, $\theta \in \left(-\frac{\pi}{2}, \frac{\pi}{2}\right)$, $\zeta \neq 0$. The additional requirement for $\Delta(\bar{x})$ to be invertible and the relevant deduction proof are given in the rest of this paper.

### III. Invertibility Analysis

**Proposition:** The $\Delta(\bar{x})$ is invertible within the region $\varphi \in \left(-\frac{\pi}{2}, \frac{\pi}{2}\right)$, $\theta \in \left(-\frac{\pi}{2}, \frac{\pi}{2}\right)$, $\zeta \neq 0$ if and only if:

$$-1 + \cos^2\theta \cdot \cos^2\phi - \cos^2\theta \cdot \cos\phi \cdot \sin\phi \neq 0 \qquad (21)$$

In this section, we present the deducing process of Equation (21) in high detail. The necessary and sufficient requirement for $\Delta(\bar{x})$ to be invertible is:

$$|\mathrm{P}^{4\times 4} \cdot \Delta(\bar{x}) \cdot \mathrm{Q}^{4\times 4}| \neq 0, \quad (\mathrm{P} \text{ and } \mathrm{Q} \text{ are invertible}) \qquad (22)$$

Before finding $\mathrm{P}$ and $\mathrm{Q}$, we start with calculating $\Delta(\bar{x})$ analytically in Equation (16). To do so, $y_1^{(4)}$, $y_2^{(4)}$, $y_3^{(4)}$, and $\ddot{y}_4$ are calculated first. The following is the deducing process of $y_1^{(4)}$.

$$\ddot{y}_1 = -\frac{1}{m} \cdot A_1 \cdot \zeta \qquad (23)$$

$$y_1^{(3)} = -\frac{1}{m} \cdot \dot{A}_1 \cdot \zeta - \frac{1}{m} \cdot A_1 \cdot \xi \qquad (24)$$

$$y_1^{(4)} = -\frac{1}{m} \cdot \ddot{A}_1 \cdot \zeta - \frac{2}{m} \cdot \dot{A}_1 \cdot \xi - \frac{1}{m} \cdot A_1 \cdot \bar{u}_1 \tag{25}$$

$y_2^{(4)}$ and $y_3^{(4)}$ are deduced in the similar way:

$$y_2^{(4)} = -\frac{1}{m} \cdot \ddot{A}_2 \cdot \zeta - \frac{2}{m} \cdot \dot{A}_2 \cdot \xi - \frac{1}{m} \cdot A_2 \cdot \bar{u}_1 \tag{26}$$

$$y_3^{(4)} = -\frac{1}{m} \cdot \ddot{A}_3 \cdot \zeta - \frac{2}{m} \cdot \dot{A}_3 \cdot \xi - \frac{1}{m} \cdot A_3 \cdot \bar{u}_1 \tag{27}$$

$\ddot{y}_4$ is deduce in Equation (28).

$$\ddot{\psi} = \left(\frac{s\phi}{c\theta} \cdot q + \frac{c\phi}{c\theta} \cdot r\right)' = A_4(\phi,\theta,p,q,r) + \frac{s\phi}{c\theta} \cdot \frac{d}{I_y} \cdot \bar{u}_3 + \frac{c\phi}{c\theta} \cdot \frac{d}{I_z} \cdot \bar{u}_4 \tag{28}$$

Notice that $A_1$, $A_2$, and $A_3$ are the functions of ($\psi$, $\theta$, $\phi$). Thus, $\dot{A}_1$, $\dot{A}_2$, and $\dot{A}_3$ are the functions of the state $\bar{x}$ in (10), containing no $\bar{u}_1$, $\bar{u}_2$, and $\bar{u}_3$. Consequently, terms $-\frac{2}{m} \cdot \dot{A}_i \cdot \xi$ in Equation (25) - (27) do not contribute to the coefficients of $\bar{u}_1$, $\bar{u}_2$, and $\bar{u}_3$.

On the other hand, $\ddot{A}_1$, $\ddot{A}_2$, and $\ddot{A}_3$ contain the calculation in differentiating the state $\bar{x}$ in (10), which generates $\bar{u}_1$, $\bar{u}_2$, and $\bar{u}_3$. Observing Equation (25) - (27), (28), calculating $\ddot{A}_1$, $\ddot{A}_2$, and $\ddot{A}_3$ is necessary to receive the coefficients of $\bar{u}_1$, $\bar{u}_2$, and $\bar{u}_3$. In the following, we are to find the terms containing $\bar{u}_1$, $\bar{u}_2$, and $\bar{u}_3$ in $\ddot{A}_1$, $\ddot{A}_2$, and $\ddot{A}_3$. We start with finding $\ddot{A}_1$.

$$\dot{A}_1 = -s\psi \cdot s\theta \cdot c\phi \cdot \dot{\psi} + c\psi \cdot c\theta \cdot c\phi \cdot \dot{\theta} - c\psi \cdot s\theta \cdot s\phi \cdot \dot{\phi} + c\psi \cdot s\phi \cdot \dot{\psi} + s\psi \cdot c\phi \cdot \dot{\phi} \tag{29}$$

Substituting $\dot{\psi}$, $\dot{\theta}$, $\dot{\varphi}$ in Equation (9), we receive Equation (30).

$$\dot{A}_1 = M_1 \cdot \left(\frac{s\phi}{c\theta} \cdot q + \frac{c\phi}{c\theta} \cdot r\right) + N_1 \cdot (p + s\phi \cdot t\theta \cdot q + c\phi \cdot t\theta \cdot r) + O_1 \cdot (c\phi \cdot q - s\phi \cdot r) \tag{30}$$

where $M_1$, $N_1$, $O_1$ are defined in (31) – (33).

$$M_1 = c\psi \cdot s\phi - s\psi \cdot s\theta \cdot s\phi \tag{31}$$

$$N_1 = s\psi \cdot c\phi - c\psi \cdot s\theta \cdot s\phi \tag{32}$$

$$O_1 = c\psi \cdot c\theta \cdot c\phi \tag{33}$$

Further, calculate the terms containing ($\dot{p}$, $\dot{q}$, $\dot{r}$) in $\ddot{A}_1$ based on the result of $\dot{A}_1$ in Equation (30). The sum of the terms containing $\dot{p}$, $\dot{q}$, $\dot{r}$, which generates $\bar{u}_2$, $\bar{u}_3$, and $\bar{u}_4$, respectively, in $\ddot{A}_1$ is illustrated in (34).

$$R_{A_1} = N_1 \cdot \dot{p} + \left(M_1 \cdot \frac{s\phi}{c\theta} + N_1 \cdot s\phi \cdot t\theta + O_1 \cdot c\phi\right) \cdot \dot{q} + \left(M_1 \cdot \frac{c\phi}{c\theta} + N_1 \cdot c\phi \cdot t\theta - O_1 \cdot s\phi\right) \cdot \dot{r} \quad (34)$$

Thus, the coefficient of $\bar{u}_1$ in Equation (25) is

$$-\frac{1}{m} \cdot A_1 \quad (35)$$

The coefficient of $\bar{u}_2$ in Equation (25) confirmed by Equation (34) is

$$-\frac{\zeta}{m} \cdot N_1 \cdot \frac{d}{I_x} \quad (36)$$

The coefficient of $\bar{u}_3$ in Equation (25) confirmed by Equation (34) is

$$-\frac{\zeta}{m} \cdot \left(M_1 \cdot \frac{s\phi}{c\theta} + N_1 \cdot s\phi \cdot t\theta + O_1 \cdot c\phi\right) \cdot \frac{d}{I_y} \quad (37)$$

The coefficient of $\bar{u}_4$ in Equation (25) confirmed by Equation (34) is

$$-\frac{\zeta}{m} \cdot \left(M_1 \cdot \frac{c\phi}{c\theta} + N_1 \cdot c\phi \cdot t\theta - O_1 \cdot s\phi\right) \cdot \frac{d}{I_z} \quad (38)$$

So far, we have received the results of the coefficients of $\bar{u}_1$, $\bar{u}_2$, $\bar{u}_3$, and $\bar{u}_4$ in $y_1^{(4)}$ in (35) – (38).

The similar procedure is conducted to calculate the coefficients of $\bar{u}_1$, $\bar{u}_2$, $\bar{u}_3$, and $\bar{u}_4$ in $y_2^{(4)}$, Equation (26). It starts with finding $\dot{A}_2$.

$$\dot{A}_2 = M_2 \cdot \left(\frac{s\phi}{c\theta} \cdot q + \frac{c\phi}{c\theta} \cdot r\right) + N_2 \cdot (p + s\phi \cdot t\theta \cdot q + c\phi \cdot t\theta \cdot r) + O_2 \cdot (c\phi \cdot q - s\phi \cdot r) \quad (39)$$

where $M_2$, $N_2$, $O_2$ are defined in (40) – (42).

$$M_2 = s\psi \cdot s\phi + c\psi \cdot s\theta \cdot s\phi \quad (40)$$

$$N_2 = -c\psi \cdot c\phi - s\psi \cdot s\theta \cdot s\phi \quad (41)$$

$$O_2 = s\psi \cdot c\theta \cdot c\phi \quad (42)$$

Further, calculate the terms containing ($\dot{p}$, $\dot{q}$, $\dot{r}$) in $\ddot{A}_2$ based on the result of $\dot{A}_2$ in Equation (39). The sum of the terms containing $\dot{p}$, $\dot{q}$, $\dot{r}$, which generates $\bar{u}_2$, $\bar{u}_3$, and $\bar{u}_4$, respectively, in $\ddot{A}_2$ is illustrated in (43).

$$R_{A_2} = N_2 \cdot \dot{p} + \left(M_2 \cdot \frac{s\phi}{c\theta} + N_2 \cdot s\phi \cdot t\theta + O_2 \cdot c\phi\right) \cdot \dot{q} + \left(M_2 \cdot \frac{c\phi}{c\theta} + N_2 \cdot c\phi \cdot t\theta - O_2 \cdot s\phi\right) \cdot \dot{r} \quad (43)$$

Thus, the coefficient of $\bar{u}_1$ in Equation (26) is

$$-\frac{1}{m} \cdot A_2 \tag{44}$$

The coefficient of $\bar{u}_2$ in Equation (26) confirmed by Equation (43) is

$$-\frac{\zeta}{m} \cdot N_2 \cdot \frac{d}{I_x} \tag{45}$$

The coefficient of $\bar{u}_3$ in Equation (26) confirmed by Equation (43) is

$$-\frac{\zeta}{m} \cdot \left(M_2 \cdot \frac{s\phi}{c\theta} + N_2 \cdot s\phi \cdot t\theta + O_2 \cdot c\phi\right) \cdot \frac{d}{I_y} \tag{46}$$

The coefficient of $\bar{u}_4$ in Equation (26) confirmed by Equation (43) is

$$-\frac{\zeta}{m} \cdot \left(M_2 \cdot \frac{c\phi}{c\theta} + N_2 \cdot c\phi \cdot t\theta - O_2 \cdot s\phi\right) \cdot \frac{d}{I_z} \tag{47}$$

The similar procedure is conducted to calculate the coefficients of $\bar{u}_1$, $\bar{u}_2$, $\bar{u}_3$, and $\bar{u}_4$ in $y_3^{(4)}$, (27).

Thus, the coefficient of $\bar{u}_1$ in Equation (27) is

$$-\frac{1}{m} \cdot A_3 \tag{48}$$

The coefficient of $\bar{u}_2$ in Equation (27) is

$$-\frac{\zeta}{m} \cdot N_3 \cdot \frac{d}{I_x} \tag{49}$$

The coefficient of $\bar{u}_3$ in Equation (27) is

$$-\frac{\zeta}{m} \cdot (N_3 \cdot s\phi \cdot t\theta + O_3 \cdot c\phi) \cdot \frac{d}{I_y} \tag{50}$$

The coefficient of $\bar{u}_4$ in Equation (27) is

$$-\frac{\zeta}{m} \cdot (N_3 \cdot c\phi \cdot t\theta - O_3 \cdot s\phi) \cdot \frac{d}{I_z} \tag{51}$$

where $N_3$, $O_3$ are defined in (52) – (53).

$$N_3 = -c\theta \cdot s\phi \tag{52}$$

$$O_3 = -s\theta \cdot c\phi \tag{53}$$

As for the coefficients of $\bar{u}_1$, $\bar{u}_2$, $\bar{u}_3$, and $\bar{u}_4$ in $\ddot{y}_4$. We can find them in Equation (28).

Thus, the coefficient of $\bar{u}_1$ in Equation (28) is

$$0 \tag{54}$$

The coefficient of $\bar{u}_2$ in Equation (28) is

$$0 \tag{55}$$

The coefficient of $\bar{u}_3$ in Equation (28) is

$$\frac{s\phi}{c\theta} \cdot \frac{d}{I_y} \tag{56}$$

The coefficient of $\bar{u}_4$ in Equation (28) is

$$\frac{c\phi}{c\theta} \cdot \frac{d}{I_z} \tag{57}$$

So far, we have found all the coefficients of $\bar{u}_1$, $\bar{u}_2$, $\bar{u}_3$, and $\bar{u}_4$ in $y_1^{(4)}$, $y_2^{(4)}$, $y_3^{(4)}$, and $\ddot{y}_4$. Thus, the decoupling matrix (Delta Matrix), $\Delta(\bar{x})$, in Equation (20) is found. That is

$$\Delta(\bar{x}) = \begin{bmatrix} \Delta_{11} & \Delta_{12} & \Delta_{13} & \Delta_{14} \\ \Delta_{21} & \Delta_{22} & \Delta_{23} & \Delta_{24} \\ \Delta_{31} & \Delta_{32} & \Delta_{33} & \Delta_{34} \\ \Delta_{41} & \Delta_{42} & \Delta_{43} & \Delta_{44} \end{bmatrix} \tag{58}$$

where

$$\Delta_{11} = -\frac{1}{m} \cdot A_1$$

$$\Delta_{21} = -\frac{1}{m} \cdot A_2$$

$$\Delta_{31} = -\frac{1}{m} \cdot A_3$$

$$\Delta_{41} = 0$$

$$\Delta_{12} = -\frac{\zeta}{m} \cdot N_1 \cdot \frac{d}{I_x}$$

$$\Delta_{22} = -\frac{\zeta}{m} \cdot N_2 \cdot \frac{d}{I_x}$$

$$\Delta_{32} = -\frac{\zeta}{m} \cdot N_3 \cdot \frac{d}{I_x}$$

$$\Delta_{42} = 0$$

$$\Delta_{13} = -\frac{\zeta}{m} \cdot \left( M_1 \cdot \frac{s\phi}{c\theta} + N_1 \cdot s\phi \cdot t\theta + O_1 \cdot c\phi \right) \cdot \frac{d}{I_y}$$

$$\Delta_{23} = -\frac{\zeta}{m} \cdot \left( M_2 \cdot \frac{s\phi}{c\theta} + N_2 \cdot s\phi \cdot t\theta + O_2 \cdot c\phi \right) \cdot \frac{d}{I_y}$$

$$\Delta_{33} = -\frac{\zeta}{m} \cdot (N_3 \cdot s\phi \cdot t\theta + O_3 \cdot c\phi) \cdot \frac{d}{I_y}$$

$$\Delta_{43} = \frac{s\phi}{c\theta} \cdot \frac{d}{I_y}$$

$$\Delta_{14} = -\frac{\zeta}{m} \cdot \left( M_1 \cdot \frac{c\phi}{c\theta} + N_1 \cdot c\phi \cdot t\theta - O_1 \cdot s\phi \right) \cdot \frac{d}{I_z}$$

$$\Delta_{24} = -\frac{\zeta}{m} \cdot \left( M_2 \cdot \frac{c\phi}{c\theta} + N_2 \cdot c\phi \cdot t\theta - O_2 \cdot s\phi \right) \cdot \frac{d}{I_z}$$

$$\Delta_{34} = -\frac{\zeta}{m} \cdot (N_3 \cdot c\phi \cdot t\theta - O_3 \cdot s\phi) \cdot \frac{d}{I_z}$$

$$\Delta_{44} = \frac{c\phi}{c\theta} \cdot \frac{d}{I_z}$$

It can be concluded easily that $\zeta$ should be nonzero if we expect delta matrix to be full rank.

The next step is to find proper invertible matrix $P^{4\times 4}$ and $Q^{4\times 4}$ in Equation (22). This step simplifies the delta matrix to the form in which we can calculate the determinant of the resultant matrix easily.

Notice that

$$P^{4\times 4} \cdot \Delta(\bar{x}) \cdot Q^{4\times 4} = \begin{bmatrix} D_{11} & s\psi \cdot c\phi & c\psi & 0 \\ D_{21} & -c\psi \cdot c\phi & s\psi & 0 \\ 0 & s\phi & s\theta & 0 \\ 0 & 0 & 0 & 1 \end{bmatrix} \tag{59}$$

where

$$D_{11} = c\psi \cdot s\theta \cdot c\phi - c\psi \cdot s\theta \cdot s\phi + s\psi \cdot s\phi + s\psi \cdot c\phi$$

$$D_{21} = s\psi \cdot s\theta \cdot c\phi - s\psi \cdot s\theta \cdot s\phi - c\psi \cdot s\phi - c\psi \cdot c\phi$$

$$P = P_1 \cdot P_2 \cdot P_3$$

$$P_1 = \begin{bmatrix} 1 & 0 & 0 & 0 \\ 0 & 1 & 0 & 0 \\ 0 & 0 & -c\theta & 0 \\ 0 & 0 & 0 & 1 \end{bmatrix}, P_2 = \begin{bmatrix} 1 & 0 & 0 & s\phi \cdot s\psi \cdot c\theta \\ 0 & 1 & 0 & s\phi \cdot s\psi \cdot c\theta \\ 0 & 0 & 1 & -s\phi \cdot s\theta \\ 0 & 0 & 0 & 1 \end{bmatrix}, P_3 = \begin{bmatrix} 1 & 0 & 0 & \frac{\zeta}{m} \cdot c\theta \\ 0 & 1 & 0 & \frac{\zeta}{m} \cdot c\theta \\ 0 & 0 & 1 & \frac{\zeta}{m} \cdot c\theta \\ 0 & 0 & 0 & -\frac{\zeta}{m} \cdot c\theta \end{bmatrix}$$

$$Q = Q_1 \cdot Q_2 \cdot Q_3$$

$$Q_1 = \begin{bmatrix} -m & 0 & 0 & 0 \\ 0 & -\frac{m \cdot I_x}{\zeta \cdot d} & 0 & 0 \\ 0 & 0 & -\frac{m \cdot I_y}{\zeta \cdot d} & 0 \\ 0 & 0 & 0 & -\frac{m \cdot I_z}{\zeta \cdot d} \end{bmatrix}, Q_2 = \begin{bmatrix} 1 & 0 & 0 & 0 \\ 0 & 1 & 0 & 0 \\ 0 & 0 & 1 & 0 \\ 0 & 0 & -\frac{s\phi}{c\phi} & 1 \end{bmatrix}, Q_3 = \begin{bmatrix} 1 & 0 & 0 & 0 \\ 1 & 1 & 0 & 0 \\ 0 & \frac{s\theta \cdot s\phi}{c\theta} & \frac{1}{c\theta} & 0 \\ 0 & 0 & 0 & \frac{1}{c\phi} \end{bmatrix}$$

Further,

$$|P^{4\times 4} \cdot \Delta(\bar{x}) \cdot Q^{4\times 4}| = -1 + \cos^2\theta \cdot \cos^2\phi - \cos^2\theta \cdot \cos\phi \cdot \sin\phi \tag{60}$$

Thus, $\Delta(\bar{x})$ is invertible if and only if the right side of (60) is nonzero, which is exactly Equation (21). The evidence of Proposition is complete.

## IV. Remarks in The Singular Zone

The singular space is determined by the following equation:

$$-1 + \cos^2\theta \cdot \cos^2\phi - \cos^2\theta \cdot \cos\phi \cdot \sin\phi = 0 \tag{61}$$

This space is defined by the variable $\phi$ and $\theta$, which are not controlled directly; the controlled variables selected in (19) do not include the roll and pitch. Consequently, avoiding the singular space defined in Equation (61) can be not straightforward or even unfeasible for some trajectories.

We visualize the singular space by defining:

$$S(\theta, \phi) = -1 + \cos^2\theta \cdot \cos^2\phi - \cos^2\theta \cdot \cos\phi \cdot \sin\phi \tag{62}$$

We plot $S(\theta, \phi)$ in Figure 1 (3D) and Figure 2 (Contour plot).

Let

$$S(\theta,\phi)=0 \tag{63}$$

And plot the result in Figure 3.

Clearly, part of this singular region is within the space of interest, $\phi\in\left(-\dfrac{\pi}{2},\dfrac{\pi}{2}\right)$, $\theta\in\left(-\dfrac{\pi}{2},\dfrac{\pi}{2}\right)$.

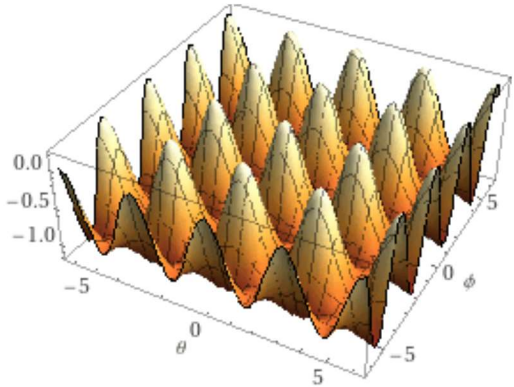

Fig. 1. 3D plot of the $S(\theta,\phi)$

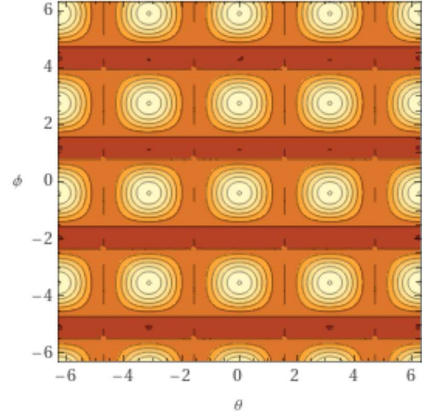

Fig. 2. Contour plot of $S(\theta,\phi)$

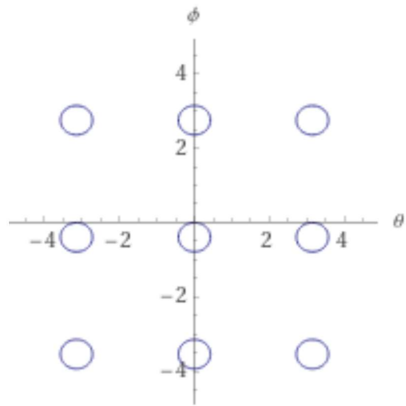

Fig. 3. $S(\theta,\phi)=0$

To avoid the singular space, the quadrotor is either outside the "circular" area in Figure 3 or inside it throughout the entire flight. Crossing the "circular" area is strictly prohibited since it causes the non-invertible problem.

This requirement is strict since we can find in Figure 3 that $(\theta,\phi)=(0,0)$ is at the singular space. While the quadrotor is near this region during most flights or even on it while hovering.

It is worth mentioning that this result is not influenced by changing the direction of the body-fixed frame; changing the positive direction of the body-fixed frame (e.g., [10]) changes the sign in several rows in delta matrix, which has no influence on the rank of a matrix.

## V. Switch Control Modification

This section puts forward an approach to avoid the above singular zone by switching to another controller when non-invertible problem occurs.

Though the previous analysis showed that yaw-position-output-based feedback linearization introduces the singular zone, attitude-altitude-output-based ($\phi, \theta, \psi, z$) feedback linearization does not [11]. Inspired from this, we replace the control rule near the singular area.

As illustrated in Figure 4, the attitude shall be lying either within the purple zone or the yellow zone during the whole flight based on the output choice $(x, y, z, \psi)$. Crossing the circular bound from the yellow zone to the purple zone is prohibited; vice versa.

To avoid crossing it, we change the controller in some area of the attitude (Figure 5). In the orange zone, we use another output choice $(\phi, \theta, \psi, z)$. While the control method for rest of the attitude area (yellow zone) remains unchanged $(x, y, z, \psi)$. Based on this, the singular condition in Figure 4 is eliminated in the whole attitude area. The attitude is allowed to change within the yellow or orange areas or crossing from one to the other.

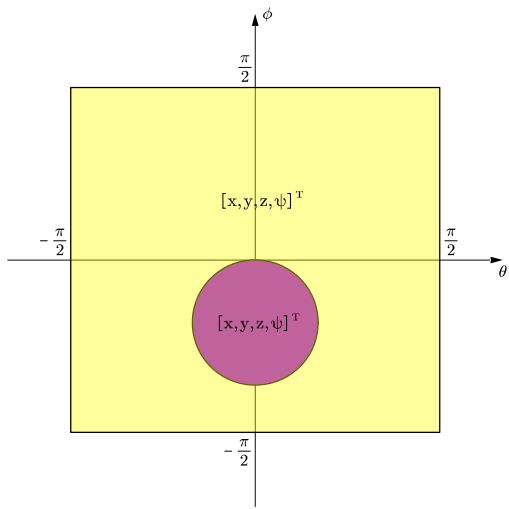 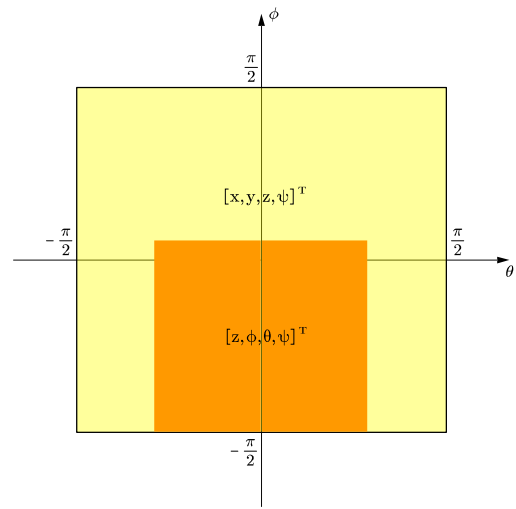

Fig. 4. $(x, y, z, \psi)$ Output Admissible Zone     Fig. 5. Switch Control Modification

The attitude zone where the output $\phi, \theta, \psi, z$ is chosen, orange zone in figure 5, is determined in Equation (64).

$$\begin{cases} -0.5 \leqslant \theta \leqslant 0.5 \\ -\frac{\pi}{2} \leqslant \phi \leqslant 0.2 \end{cases} \tag{64}$$

The whole Simulink block is pictured in Figure 6. The relevant parts are signed.

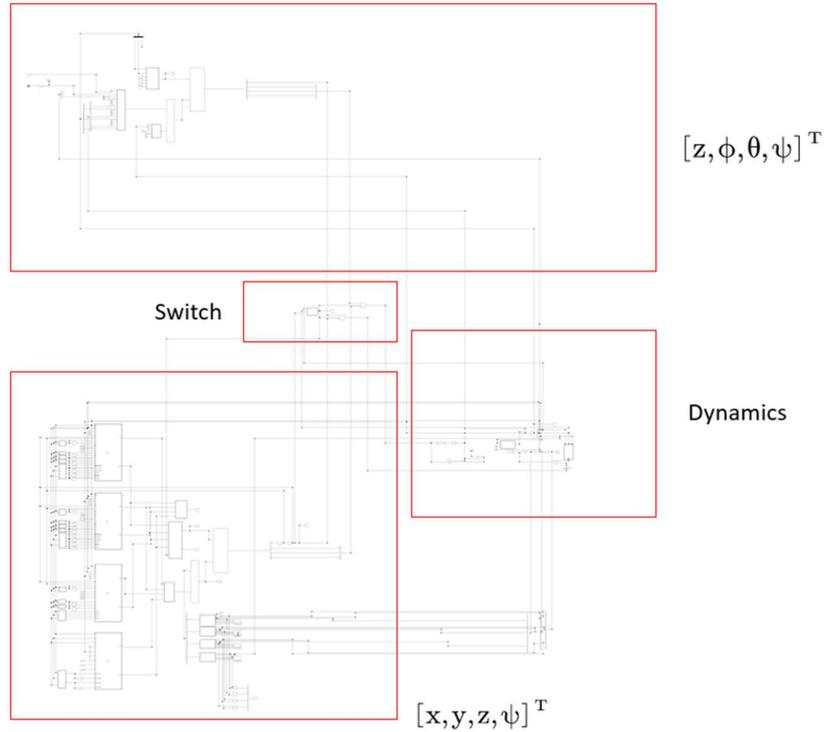

Fig.6. The Simulink diagram for the modified controller equipped system

The initial attitude is set as $(\phi_0, \theta_0, \psi_0) = (0.5, 0.5, 0)$.

In the first experiment, the position reference and the attitude reference for two controllers are marked in Figure 7. Figure 8 shows the result of the attitude trajectory. It can be seen that attitude $(\phi, \theta)$ starts from $(0.5, 0.5)$ under the dominance of the yaw-position controller (yellow zone). After some time, it enters the orange zone governed by altitude-attitude controller and is captured by this zone before stabilizing at $(\phi, \theta) = (0.01, 0)$.

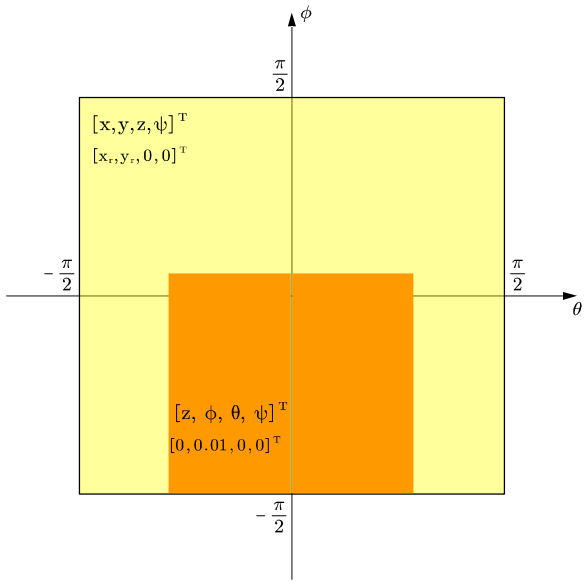
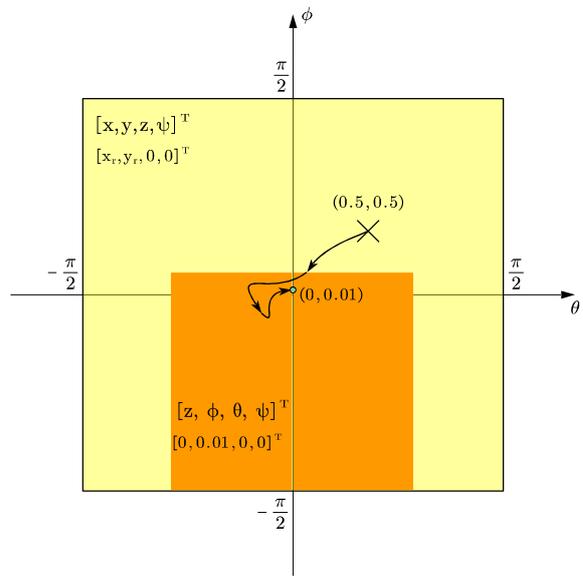

Fig.7. The references for the two controllers    Fig. 8. Attitude trajectory

In the second experiment, we would like to stabilize our attitude at $(\phi_f, \theta_f) = (0.5, -0.5)$. Thus, we set the reference for both controllers in Figure 9. Figure 10 shows the result of the attitude trajectory.

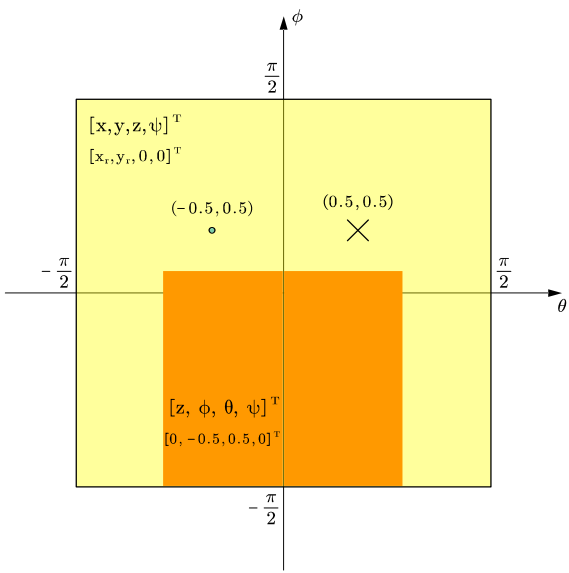
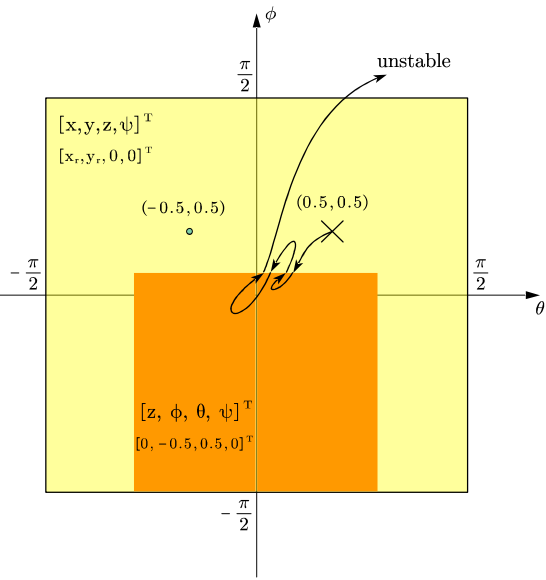

Fig.9. The references for the two controllers    Fig. 10. Attitude trajectory

The attitude witnessed several switches in controllers before being unstable finally. Although the altitude-attitude control zone (orange zone) tries to let the attitude reach $(\phi_f, \theta_f) = (0.5, -0.5)$, altitude-attitude controller is switched to the yaw-position controller once the attitude escapes the orange zone.

The yaw-position controller has no effect to drive the attitude, especially roll and pitch, $\varphi$ and $\theta$. Thus the attitude in the yellow zone can escape the area of interest, $\phi \in \left(-\frac{\pi}{2}, \frac{\pi}{2}\right)$, $\theta \in \left(-\frac{\pi}{2}, \frac{\pi}{2}\right)$, becoming unstable. The attitude signal history is also plotted in Figure 11.

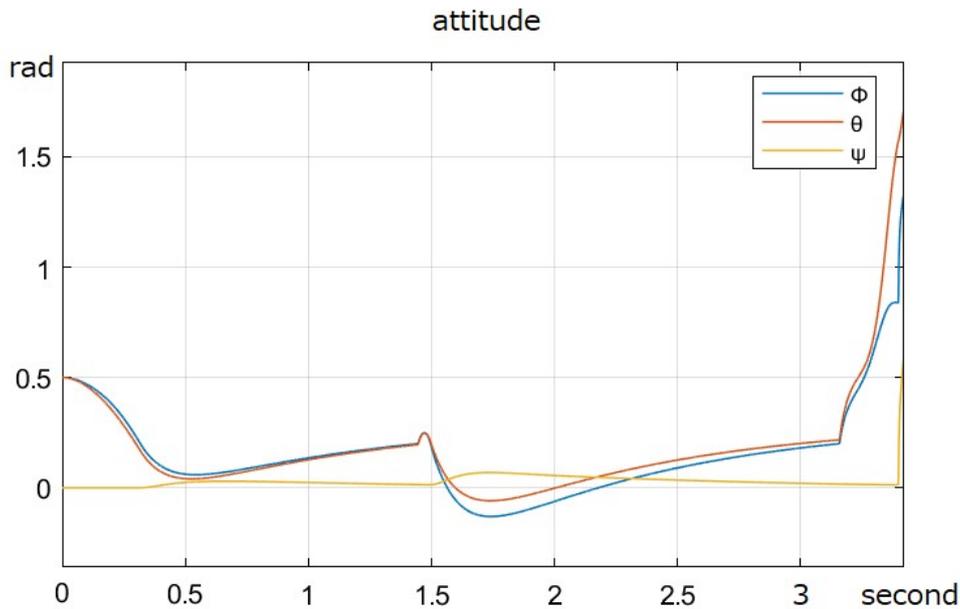

Fig. 11. Attitude history before being unstable

### VI. Conclusion and Discussion

The decoupling matrix is not invertible for the yaw-position output combination. Thus, the relevant controller risks encountering the singular problem which causes the failure in controlling.

Substituting part of the yaw-position controller by altitude-attitude controller can avoid non-invertible problem. The final stabilizing attitude lies inside the region governed by the altitude-attitude controller. Attempts in driving the attitude to a target inside the yaw-position controller dominated attitude region can cause unstable and is hard to realize.

We did not successfully stabilize the attitude in the attitude zone controlled by yaw-position controller.

One way to avoid the singularity while maintaining the largest number of controlled variables (4) is to pick other controlled variable combinations. E.g., pick attitude and altitude ($\phi, \theta, \psi, z$) as controlled variables [11]. The delta matrix in [11] is strictly full rank within the range of interest.

Another way is to sacrifice the number of the controlled variables. For example, [12] use a cascade control structure with three controlled variables for each layer to control all the 6 variables subsequently. The singularity problem is also avoided in this way.

However, a control method which directly avoiding this singular space is still an open question.